\documentstyle[12pt]{article}
\begin{document}
\vspace{0.5 cm}
\begin{center}
{\large \bf {\large Multifragment production in Au+Au at 35 MeV/u}}
\end{center}
\vspace{0.5 cm}
\small
{M. D'Agostino$^a$, P. F. Mastinu$^a$, P. M. Milazzo$^a$, M. Bruno$^a$,
D. R. Bowman$^g$, P. Buttazzo$^b$, L. Celano$^c$, N. Colonna$^c$,
J. D. Dinius$^f$, A. Ferrero$^{d,h}$, M. L. Fiandri$^a$, C. K. Gelbke$^f$,
T. Glasmacher$^f$, F. Gramegna$^e$, D. O. Handzy$^f$, D. Horn$^g$,
W. C. Hsi$^f$, M. Huang$^f$, I. Iori$^d$, G. J. Kunde$^f$, M. A. Lisa$^f$,
W. G. Lynch$^f$, L. Man\-du\-ci$^a$, G. V. Margagliotti$^b$,
C. P. Montoya$^f$,
A. Moroni$^d$, G. F. Peaslee$^f$, F. Petruzzelli$^d$, L. Phair$^f$,
R. Rui$^b$,
C. Schwarz$^f$, M. B. Tsang$^f$, G. Vannini$^b$, C. Williams$^f$}

\vspace{0.2 cm}
 {\it Multics / Miniball collaboration}
\vspace{0.2 cm}

\scriptsize

$^{a}$ Dipartimento di Fisica and INFN, Bologna, Italy

$^{b}$ Dipartimento di Fisica and INFN, Trieste, Italy

$^{c}$  INFN, Bari, Italy

$^{d}$ Dipartimento di Fisica and INFN, Milano, Italy

$^{e}$ INFN, Laboratori Nazionali di Legnaro, Italy

$^{f}$ NSCL, Michigan State University, USA

$^{g}$ Chalk River Laboratories, Canada

$^{h}$ C. N. E. A. Buenos Ayres, Argentina
\vspace{0.5 cm}

\small
Multifragment disintegration has been measured with a high efficiency
detection system for the reaction $Au + Au$ at $E/A = 35\ MeV$.
{}From the event shape analysis and the comparison with the predictions of a
many-body trajectories calculation the data, for central collisions,
are compatible with a fast emission from a unique fragment source.
\vspace{0.5 cm}

\normalsize
\rm
The disassembly of highly excited systems remains an open problem in the
investigation of intermediate energy Nucleus-Nucleus
collisions~\cite{general,theo}.
One of the challenging questions for head-on collisions is whether light
particles and fragments emission is compatible with the fast emission from a
unique thermalized source or it can still be explained in the
deep-inelastic framework.

Several recent experimental studies of central collisions, performed
with very heavy nuclei at different incident energies, give different
indications on this point~\cite{100mev,neutron,lecolley,indrabormio}.
In $100\ MeV/u$ $Au + Au$~\cite{100mev} central collisions, dynamical and
statistical analyses~\cite{smm100} suggest that the large multiplicities,
observed for light particles and Intermediate Mass Fragments, are compatible
with the prompt multifragmentation of a heavy, thermalized composite system
with freeze-out density $\approx \frac{1}{3} \div \frac{1}{6}$ of the
normal nuclear density ($\rho_0 = 0.15\ fm^{-3}$), even if the fragment
probability emission resulted strongly influenced by the radial
flow~\cite{100mev,poggi}.
In the nearly symmetric $Pb+Au$ reaction at $29 MeV/u$~\cite{neutron} the
charged products emission, studied for increasing neutron multiplicity,
shows that the emission of Intermediate Mass Fragments becomes the largest
component of the cross section at the expense of Projectile Like Fragments
and Fission Fragments emission.
On the other hand an analysis, mainly based on the event shape, of
the same reaction $Pb+Au$~\cite{lecolley,durand} at the same incident energy
seems to reveal that, even selecting the most central collisions,
the largest part of the total reaction cross section is due to strongly
damped
binary collisions.
Other studies on light particle and fragment
emission~\cite{indrabormio} seem to confirm that, even
selecting the most central collisions, the Incomplete Fusion cross section
vanishes when reactions involving heavy projectile and targets at incident
energies greater than $\approx 35 MeV/u$ are studied.

In this Letter we report the results of the analysis of central collisions
for
the reaction $Au + Au$ at $35\ MeV/u$, measured with a high efficiency
detection
system.  We will show that the observed fragment emission is compatible
with the fast emission from a unique equilibrated intermediate nuclear
system.

The experiment was performed at the National Superconducting Cyclotron
Laboratory of the Michigan State University.
Beams of Au ions at $E/A=35$ MeV incident energy, accelerated by the K1200
cyclotron, were used to bombard Au foils of approximately 5 mg/cm$^2$ areal
density.
Light charged particles and fragments with charge $Z \le 20$ were detected
at
$23^o \le \theta_{lab} \le 160^o$ by 159 phoswich detector elements of the
MSU {\it Miniball}~\cite{mini}.
Reaction products with charge $Z \le 83$ were detected at
$3^o \le \theta_{lab} < 23^o$ by the {\it Multics}
array~\cite{strum}.
The charge identification thresholds were about 2, 3, 4 MeV/u
in the {\it Miniball} for Z= 3, 10, 18, respectively and about 1.5 MeV/u
in the
{\it Multics} array independent of the fragment charge.
The geometric acceptance of the combined apparata was greater than
87\% of $4\pi$.

{}From the total charged particle multiplicity $N_c$ the reduced impact
parameter
$\hat{b}$ was determined, following Ref.\cite{tsang}.
Additional high statistics measurements were done using a shield, covering
$\theta_{lab} < 8^o$ in order to minimize the radiation damage of the most
forward detectors~\cite{footnote}.
In this way more than $10^5$ events were collected for a centrality cut
$N_c>24, \hat{b}\le 0.3$, which selects 10\% of the total measured reaction
cross section.
In this range of impact parameters the measured light particles
($Z \le 2$) multiplicity $M_l$ has a mean value $\sim 20$ and the
fragment ($Z >2$) multiplicity $N_f$ distribution has a gaussian shape with
mean value $5.6$ and standard deviation $1.8$.
The mean value of $M_l$ is very much higher than the one chosen in
Ref.\cite{lecolley} ($M_l >10$) to identify the central collisions.
The requirement $M_l > 10$, corresponding to
$N_c > 15$, would select an impact parameter range $\hat{b}\le 0.6$.
Moreover the gaussian $N_f$ distribution looks very different from the
distribution shown in Fig.~2 of Ref.\cite{lecolley}, where the two-fragments
events represent the largest part of the measured cross section.

To investigate the fragment emission patterns we first performed a shape
analysis, looking at the sphericity, coplanarity and flow angle,
variables sensitive to the dynamics of the fragmentation
process~\cite{cugnon}.
The emission of fragments from a unique source should be on the average
isotropic in momentum space and the event shape should fluctuate
around a sphere.
Conversely in peripheral reactions the forward-backward emission of
fragments
from the spectator-like sources should lead to an event shape elongated
along
the beam axis, to a decrease of the sphericity value and to flow angles
peaked in the forward direction.

In this analysis only events satisfying the constraint that at least
70\% of the incoming momentum had been detected, were considered.
For central and intermediate impact parameters, where the particle and
fragments detection is less influenced than in peripheral collisions by the
energy thresholds and the angular acceptance, a further constraint on the
total detected charge (70\% of the total charge) was applied.

The momentum tensor has been evaluated:
$$T_{ij} = \Sigma \frac {p_i^{(n)} \cdot p_j^{(n)}}{p^{(n)}}
\hspace{0.5 cm} (i, j = 1, 2, 3) \eqno(1)$$
where $p_i^{(n)}, p_j^{(n)}$ are the $i-th$ and $j-th$ cartesian projections
of the momentum $\vec {p}^{(n)}$ of the $n-th$ fragment in the centre
of mass frame.
The sum runs over the number of fragments ($Z>2$) detected in each event.
The diagonalization of the flow tensor gives three eigen-values $\lambda_i$
and three eigen-vectors $\vec{e_i}$. The event shape is an oriented
ellipsoid
with the principal axes parallel to the eigen-vectors.
The sphericity and coplanarity variables are, respectively, defined as:
$$ S = 1.5 \cdot (1 - \lambda_1), \hspace{0.5 cm}
 C = \frac {\sqrt 3} {2} \cdot (\lambda_2 - \lambda_3) \eqno(2)$$
where $\lambda_1, \lambda_2, \lambda_3$ are the ordered eigenvalues
($\lambda_1 \ge \lambda_2 \ge \lambda_3$), normalized to their sum.
The flow angle $\theta_{flow}$ is the angle between the eigenvector
$\vec{e_1}$ for the largest eigenvalue $\lambda_1$ and the beam axis:
$$ cos(\theta_{flow}) = \vec{e_1}\cdot \hat{k} \eqno(3)$$
Events with more than two detected fragments were used in the event shape
analysis: two-body events, indeed, correspond mainly to peripheral
reactions and do not give significant information, being two of the three
eigenvalues zero for the momentum conservation.

In Fig.~1 the $S-C$ plot is shown, together with the cosine
of the flow angle for three different gates on $N_c$.
For $N_c<15$ which corresponds to $\hat{b}>0.6$, as expected
for peripheral events, the memory of the entrance channel dominates: one or
two of the three eigenvalues extracted from the momentum tensor are
nearly 0,
so that the event is {\it pencil} or {\it disk} shaped and
$cos(\theta_{flow})$ is peaked in the forward direction.
With a rough constraint on the centrality, i.e. requiring $N_c$
larger than 15 ($\hat{b}\le 0.6$), the events do not show a well defined
shape, though the centroid of the events is shifted, with respect to the
more
peripheral collisions, towards the corner which represents spherical events.
This behaviour reflects in the $cos(\theta_{flow})$ distribution, which is
less forward peaked than in the previous case.
A more stringent constraint on the centrality ($N_c > 24\ ,\hat{b}\le 0.3$)
leads to a drastic change of the $S-C$ pattern: in these collisions the
eigenvalues are very similar to each other, so that mainly events with shape
close to a sphere are clearly present and $cos(\theta_{flow})$ is randomly
distributed, as expected in the case of fragments emitted from a unique
source.
Taking into account the modifications of the event shape with the
requirement on
the centrality and considering that the applied criterion on the total
detected parallel momentum and the total charge does not eliminate the heavy
residues, if they exist, we can deduce that the contribution from deep
inelastic
reactions is negligible at such small impact parameters~\cite{bormiom}.

The next step of the shape analysis consisted in the comparison of the
experimental mean values $<S>$ and $<C>$ and their standard
deviations $\Delta S$, $\Delta C$ with the prediction of a many-body
trajectory
calculation~\cite{thomas}, which has as basic assumption the uniqueness of
an emitting system with zero angular momentum.

The experimental data considered in the following were selected with
the centrality cut $N_c > 24$ ($\hat{b} \le 0.3$).
In the simulation both charge and energy of the reaction products are
selected
by randomly sampling the experimental single-particle yield for this cut.
The fragments are emitted from a spherical source of radius $R_s$ and charge
$Z_s$.
The individual emission time for each fragment is assumed to follow an
exponential probability distribution, characterized by a decay constant
$\tau$.
The emission is isotropic in the reference frame of the emitting system.
A possible collective radial expansion can be accounted for by a further
parameter $v_{coll}$, which allows to increase the fragment velocity by
a component $\vec{v} (\vec{r}) = v_{coll} \frac{\vec r}{R_s}$, which
attains its maximum value at the surface $R_s$ of the source.
The simulated events were treated in the same way as the experimental data,
after filtering~\cite{effic} them with the geometrical acceptance,
granularity,
energy thresholds and finite energy and angular resolutions.
In our case, due to the mass symmetry in the entrance channel, the fragment
source is at rest in the centre of mass frame.
This give the advantage that no hypothesis is needed on the source velocity
(on the percentage of the momentum transfer) contrary to the case of
asymmetric reactions~\cite{daniel}, when calculating the laboratory fragment
velocities.

The comparison between $<S>, <C>$ is significant and represents a check
of the
compatibility between the experimental observables and the decay of a unique
source, provided that a set of input parameters ($Z_s, R_s, \tau, v_{coll}$)
is found, reproducing the fragment momenta used in equation $(1)$ to build
the
momentum tensor.
To this aim we performed several calculations, varying in a wide range the
input parameters. For each set the predicted fragment emission velocities
as a function of the emission angle and the fragment reduced velocity
correlation functions have been compared to the experimental data.
The experimental $N_c$, $N_f$ and $Z_{bound}$ (total charge bound in form
of fragments with $Z>2$) spectra were continuously used to keep under
control
the reasonability of the predictions.

We found that a source with charge $\approx 86\%$ of the total charge
($Z_s = 138$), freeze-out density $\sim \rho_0/4$ (radius $R_s = 13 fm$),
which
emits the fragments with an average time between successive fragment
emissions~\cite{durand} $\tau \approx 85 fm/c$ reproduces the previously
mentioned observables, provided that an expansion radial velocity
$\approx 1.4 cm/ns$ is taken into account.
As can be seen in Figs.~2~a), ~b) and ~c) the experimental distributions of
the fragment emission velocities as a function of the emission angle
$\theta_{cm}$ are very well reproduced, irrespective of the selected
fragment
charge, even at forward and backward centre of mass angles, most affected by
the experimental acceptance.

Since the isotropy of the fragment emission, assumed by the calculation,
implies that the emission velocity, for a given fragment charge, is
constant over the whole range of $\theta_{cm}$, we investigated more in
detail the main reasons of the rise of the experimental distribution
at small centre of mass angles and of the dip at backward $\theta_{cm}$.
It is important indeed to understand whether these distortions
can be ascribed to the boost of sources not at rest in the centre of mass
or they are only due to experimental limitations.
A simple kinematical calculation showed that, starting from
a constant emission velocity with gaussian profile (in the calculation of
Fig.~2~d) $3 cm/ns$ with standard deviation $1 cm/ns$),
the combined effects of the laboratory angular and velocity acceptance
do not sharply cut the spectrum but they select at the most forward
(backward)
angles the highest (lowest) values of the velocity distribution.
In particular the enhancement at $\theta_{cm}< 50^o$ is mainly due to the
forward angular limitation and the dip at $\theta_{cm} > 130^o$ to the
velocity thresholds.
For $\theta_{cm} \approx 50^o \div 130^o$ the distribution is only slightly
affected by the experimental limitations, so that the mean value in
this angular range corresponds to the {\it true} emission velocity.
For sake of comparison we report in Fig.~2~d) the experimental emission
velocity for fragments with $Z = 6$, which have in the $\theta_{cm}$ range
$50^o \div 130^o$ a mean emission velocity $\sim 3 cm/ns$.

The reproduction of the relative fragment momenta was then checked
through the comparison of the two-fragments correlation functions.
The shape of the correlation function at small values of the reduced
velocities
is a measure of the spatial separation of the emitted fragments and is
therefore sensitive to the input parameters of the calculation.

The correlation functions~\cite{trockel,al} were calculated by:
$$1+ R = C \frac{Y(v_{red})} {Y_{mix}(v_{red})} \eqno(4)$$
where $v_{red}$ is the reduced velocity of fragments $i$ and
$j \ (i \not= j)$
(charges $Z_i$ and $Z_j$):
$$v_{red}  = \frac{ \mid \vec v_i - \vec v_j \mid} {\sqrt (Z_{i}+Z_j)}
\eqno(5)$$
$Y(v_{red})$ and $Y_{mix}(v_{red})$ are the coincidence and mixed yields for
fragment pairs of reduced velocity $v_{red}$.
The mixed yield was constructed by means of the mixing event technique,
$C$ is a normalization factor fixed by the requirement to have the
same number of true and mixed pairs~\cite{trockel}.

We analyzed separately fragments detected in the {\it Multics} array and
in the
{\it Miniball}, since the solid angle covered by the apparata is very
different and an average on the whole solid angle would lead to a loss of
information.
{}From Figs.3~a),~b) (fragments detected in {\it Multics}) and Figs.3~c),~d)
(fragments detected in the {\it Miniball}) one can see that the many-body
trajectories code well reproduces the experimental correlation functions,
irrespective of the selected fragment charge, assuming an average time
between
successive fragment emissions $\tau = 85\ fm/c$
and a collective radial expansion $v_{coll}\approx 1.4 cm/ns$.
Varying $\tau$ by some tens of $fm/c$ the experimental correlation functions
are not so well reproduced: the decreased or increased distance among the
emitted fragments introduces additional correlations or anticorrelations,
respectively, not present in the data~\cite{schapiro}.
It has to be noted that the experimentally observed {\it Coulomb hole}
at small values of the reduced velocity leads to the same choice of the
parameter $\tau$ both in the case of fragments emitted at small relative
angles
(detected in {\it Multics}) and in the case of fragments emitted at large
relative angles (detected in the {\it Miniball}).
In addition in the first case the selection of $\tau$ can be performed
by checking even the reproduction of the enhancement of the correlation
function at $v_{red}\approx 20$. This bump, due to the mutual Coulomb
repulsion between the closely emitted partners (see Fig.s~3~a) and b)), is
sensitive to the increase of the proximity of the fragments induced by the
decrease of $\tau$.

The small value found for the radial collective velocity, although
consistent with the extrapolation to $35 MeV/u$ of data at higher
energies~\cite{hirshegg}, should be thoroughly investigated before making
any
conclusion.
It could be ascribed either to the assumption of non overlapping fragments
or to the schematic treatment of the Coulomb component: a change of the
emission geometry, which reflects in an increase of the fragment Coulomb
energy, could compensate the need of a radial expansion.
Furthermore in a recent theoretical work~\cite{largemom} on the
effect of the angular momentum on the statistical fragmentation, it was
found
that a rotation mechanism could explain some features previously
ascribed to a collective flow of the nuclear matter.
The comparison to predictions of statistical models which take into account
either the radial flow or the angular momentum for systems with mass of some
hundreds of nucleons should be performed, but this is beyond the aim of this
Letter.

Since the observed fragment emission patterns are well reproduced by the
many-body trajectory calculation, the comparison between experimental and
predicted $<S>, <C>$ and their standard deviations $\Delta S$, $\Delta C$
as a function of the fragment multiplicity becomes significant to
draw a conclusion on the compatibility of our data with the uniqueness of
the decaying system.
{}From Fig.~4 the very good agreement between data and predictions is evident,
not only for $<S>, <C>$, but even for the standard deviations
$\Delta S, \Delta
C$, irrespective of the selected fragment multiplicity.
Observing the unfiltered predictions (dashed lines in Fig.~4), we
can deduce that the effects of the experimental inefficiencies very slightly
decrease $<S>$ and increase $<C>$.
Moreover events with $N_f=3$, not present in the unfiltered predictions, are
only few percent both in the experimental data and in the filtered
predictions.
We would like to stress that this agreement is not a trivial consequence of
the reproduction of the fragment multiplicity.
Indeed even in the case of peripheral collisions (Fig.~1~a)) events with
fragment multiplicity up to 7 were measured, but in this case the
forward-backward emission flattens the fragment momenta into a plane and
$<S>$ and $<C>$ reveal {\it pencil/disk} shaped events.

In conclusion, for the central collisions of the reaction $Au + Au$ at
$E/A = 35 \ MeV$, the good agreement among the measured observables and the
predictions of a many-body trajectories code confirms both the assumptions
on the uniqueness of the fragment source and on the isotropy of the fragment
emission.

\vspace{0.5 cm}
This work has been supported in part by funds of the Italian Ministry
of University and Scientific Research.
The technical assistance of R. Bassini, C. Boiano, S. Brambilla,
G. Busacchi,
A. Cortesi, M. Malatesta and R. Scardaoni during the measurements is
gratefully
acknowledged.

\newpage
{\bf {\Large {Figure captions}}}\\
\small
{\bf Fig.~1:} { Experimental Sphericity- Coplanarity linear contour plot and
cos($\theta_{flow}$). Upper panels: $N_c<15$ ($\hat{b}>0.6$), intermediate
panels: $N_c\ge 15$ ($\hat{b}\le 0.6$), lower panels: $N_c > 24$
($\hat{b}\le 0.3$).}\\
{\bf Fig.~2:} {~a), ~b), ~c): Experimental emission velocities (points) for
fragments with charge 4, 8 and 10, compared with the many-body trajectory
calculations (line). The statistical experimental error is smaller than
the size of the points.\\
{}~d) Experimental emission velocities for fragments with charge 6 (open
points)
compared with a constant emission velocity ($3 cm/ns$ with standard
deviation
$1 cm/ns$) (line) filtered with the constraints
$8^o \le \theta_{lab} \le 160^o$,
$v_{threshold} = 1.5 cm/ns\ (\theta_{lab}<23^o) ,v_{threshold} = 2.5 cm/ns\
(\theta_{lab}\ge23^o)$.}\\
{\bf Fig.~3:} { Two fragment correlation functions for:
a) $3 \le Z_{IMF} \le 30$ and $8^o \le \theta_{lab} < 23^o$,
b) $3 \le Z_{IMF} \le 20$ and $8^o \le \theta_{lab} < 23^o$,
c) $3 \le Z_{IMF} \le 20$ and $23^o \le \theta_{lab} \le 160^o$,
d) $3 \le Z_{IMF} \le 10$ and $23^o \le \theta_{lab} \le 160^o$.
Open points show experimental data. The solid, dashed and dotted lines are
the
the many-body trajectory predictions for $\tau = 85 ,50, 150 fm/c$
respectively.}\\
{\bf Fig.~4:} { Mean sphericity $<S>$ and coplanarity $<C>$ and their
standard deviations $\Delta S$, $\Delta C$ as a function of $N_f$.
Experimental $<S>, <C>, \Delta S, \Delta C$ are reported as full symbols and
vertical bars, respectively.
$<S>, <C>$ calculated from the filtered predictions are reported as open
symbols and $<S> \pm \Delta S, <C> \pm \Delta C$ as full lines.
$<S> \pm \Delta S, <C> \pm \Delta C$ from the unfiltered predictions are
shown
by dashed lines.}\\

\begin{thebibliography}{99}
\bibitem{footnote} The shield was thick enough to stop the scattered $Au$
ions.
\end{thebibliography}
\end{document}